\documentclass[12pt]{article}
\usepackage{amsfonts}
\usepackage{amssymb}
\usepackage{graphics,amsmath}

\usepackage{color}

\def\hybrid{\topmargin -20pt    \oddsidemargin 0pt
        \headheight 0pt \headsep 0pt
        \textwidth 6.35in       
        \textheight 9.25in       
        \marginparwidth .875in
        \parskip 5pt plus 1pt   \jot = 1.5ex}

\hybrid

\def\baselinestretch{1.2}

\catcode`\@=11

\def\marginnote#1{}
\newcount\hour
\newcount\minute
\newtoks\amorpm
\hour=\time\divide\hour by60
\minute=\time{\multiply\hour by60 \global\advance\minute by-\hour}
\edef\standardtime{{\ifnum\hour<12 \global\amorpm={am}%
        \else\global\amorpm={pm}\advance\hour by-12 \fi
        \ifnum\hour=0 \hour=12 \fi
        \number\hour:\ifnum\minute<10 0\fi\number\minute\the\amorpm}}
\edef\militarytime{\number\hour:\ifnum\minute<10 0\fi\number\minute}

\def\draftlabel#1{{\@bsphack\if@filesw {\let\thepage\relax
   \xdef\@gtempa{\write\@auxout{\string
      \newlabel{#1}{{\@currentlabel}{\thepage}}}}}\@gtempa
   \if@nobreak \ifvmode\nobreak\fi\fi\fi\@esphack}
        \gdef\@eqnlabel{#1}}
\def\@eqnlabel{}
\def\@vacuum{}
\def\draftmarginnote#1{\marginpar{\raggedright\scriptsize\tt#1}}

\def\draft{\oddsidemargin -.5truein
        \def\@oddfoot{\sl preliminary draft \hfil
        \rm\thepage\hfil\sl\today\quad\militarytime}
        \let\@evenfoot\@oddfoot \overfullrule 3pt
        \let\label=\draftlabel
        \let\marginnote=\draftmarginnote
   \def\@eqnnum{(\theequation)\rlap{\kern\marginparsep\tt\@eqnlabel}%
\global\let\@eqnlabel\@vacuum}  }

\def\preprint{\twocolumn\sloppy\flushbottom\parindent 2em
        \leftmargini 2em\leftmarginv .5em\leftmarginvi .5em
        \oddsidemargin -.5in    \evensidemargin -.5in
        \columnsep .4in \footheight 0pt
        \textwidth 10.in        \topmargin  -.4in
        \headheight 12pt \topskip .4in
        \textheight 6.9in \footskip 0pt
        \def\@oddhead{\thepage\hfil\addtocounter{page}{1}\thepage}
        \let\@evenhead\@oddhead \def\@oddfoot{} \def\@evenfoot{} }

\def\numberbysection{\@addtoreset{equation}{section}
        \def\theequation{\thesection.\arabic{equation}}}

\def\underline#1{\relax\ifmmode\@@underline#1\else
        $\@@underline{\hbox{#1}}$\relax\fi}

\def\titlepage{\@restonecolfalse\if@twocolumn\@restonecoltrue\onecolumn
     \else \newpage \fi \thispagestyle{empty}\c@page\z@
        \def\thefootnote{\fnsymbol{footnote}} }

\def\endtitlepage{\if@restonecol\twocolumn \else \newpage \fi
        \def\thefootnote{\arabic{footnote}}
        \setcounter{footnote}{0}}  

\catcode`@=12
\relax

\def\figcap{\section*{Figure Captions\markboth
        {FIGURECAPTIONS}{FIGURECAPTIONS}}\list
        {Figure \arabic{enumi}:\hfill}{\settowidth\labelwidth{Figure
999:}
        \leftmargin\labelwidth
        \advance\leftmargin\labelsep\usecounter{enumi}}}
 \relax
\def\tablecap{\section*{Table Captions\markboth
        {TABLECAPTIONS}{TABLECAPTIONS}}\list
        {Table \arabic{enumi}:\hfill}{\settowidth\labelwidth{Table
999:}
        \leftmargin\labelwidth
        \advance\leftmargin\labelsep\usecounter{enumi}}}
 \relax
\def\reflist{\section*{References\markboth
        {REFLIST}{REFLIST}}\list
        {[\arabic{enumi}]\hfill}{\settowidth\labelwidth{[999]}
        \leftmargin\labelwidth
        \advance\leftmargin\labelsep\usecounter{enumi}}}
 \relax
\makeatletter
\newcounter{pubctr}
\def\publist{\@ifnextchar[{\@publist}{\@@publist}}
\def\@publist[#1]{\list
        {[\arabic{pubctr}]\hfill}{\settowidth\labelwidth{[999]}
        \leftmargin\labelwidth
        \advance\leftmargin\labelsep
        \@nmbrlisttrue\def\@listctr{pubctr}
        \setcounter{pubctr}{#1}\addtocounter{pubctr}{-1}}}
\def\@@publist{\list
        {[\arabic{pubctr}]\hfill}{\settowidth\labelwidth{[999]}
        \leftmargin\labelwidth
        \advance\leftmargin\labelsep
        \@nmbrlisttrue\def\@listctr{pubctr}}}
 \relax
\makeatother
\newskip\humongous \humongous=0pt plus 1000pt minus 1000pt

\newif\ifdtup

\relax

\def\be{\begin{equation}}
\def\ee{\end{equation}}
\def\ba{\begin{eqnarray}}
\def\ea{\end{eqnarray}}

\DeclareMathOperator\sech{sech}

\DeclareMathOperator\arcsinh{arcsinh}

\def\a{\alpha}

\def\no{\noindent}

\def\IR{\relax{\rm I\kern-.18em R}}
\def\II{\relax{\rm 1\kern-.35em1}}

\renewcommand{\theequation}{\thesection.\arabic{equation}}
\csname @addtoreset\endcsname{equation}{section}

\def\IR{\relax{\rm I\kern-.18em R}}
\def\inv{^{\raise.15ex\hbox{${\scriptscriptstyle -}$}\kern-.05em 1}}

\begin{document}

\begin{titlepage}
\begin{center}

\vskip .5in

{\LARGE Spinning strings in the $\eta$-deformed Neumann-Rosochatius system}
\vskip 0.4in

{\bf Rafael Hern\'andez} \phantom{x} and \phantom{x}
{\bf Juan Miguel Nieto} 
\vskip 0.1in

Departamento de F\'{\i}sica Te\'orica I \\
Universidad Complutense de Madrid \\
$28040$ Madrid, Spain \\
{\footnotesize{\tt rafael.hernandez@fis.ucm.es, juanieto@ucm.es}}

\end{center}

\vskip .4in

\centerline{\bf Abstract}
\vskip .1in
\no
The sigma-model of closed strings spinning in the $\eta$-deformation of $AdS_{5} \times S^{5}$ leads to an integrable deformation of the one-dimensional 
Neumann-Rosochatius mechanical system. In this article we construct general solutions to this system that can be written in terms of elliptic functions. 
The solutions correspond to closed strings with non-constant radii rotating with two different angular momenta in an $\eta$-deformed three-sphere. We analyse the reduction 
of the elliptic solutions for some limiting values of the deformation parameter. For the case of solutions with constant radii we find the 
dependence of the classical energy of the string on the angular momenta as an expansion in the 't Hooft coupling.  
\noindent

\vskip .4in
\noindent

\end{titlepage}

\vfill
\eject

\def\baselinestretch{1.2}


\baselineskip 20pt


\section{Introduction}

The identification of the sigma-model of bosonic strings spinning in $AdS_5 \times S^{5}$ with the Neumann and the Neumann-Rosochatius systems 
was one of the first steps towards the uncovering of the integrable structure underlying the AdS/CFT correspondence \cite{NR}. The Neumann system 
is an integrable model of harmonic oscillators restricted to move on a sphere. The case of the Neumann-Rosochatius system includes an 
additional centrifugal barrier term. The equivalence of the spinning string ansatz to the Neumann and the Neumann-Rosochatius integrable systems 
allowed a beautiful description of quite general string configurations in terms of solutions to these mechanical models and proved useful 
to find their energies as functions of their spins and angular momenta, which lead to very precise comparisons with the corresponding 
gauge theory duals (see \cite{Treview} for a review). 

A natural problem is the extension of the spinning string ansatz to the study of strings rotating in less symmetric backgrounds that preserve integrability. 
The integrable deformations of the sigma-model of type IIB strings on $AdS_5 \times S^{5}$ can be divided in three different classes, 
referred to as $\eta$-deformations \cite{eta}, $\lambda$-deformations \cite{lambda} and deformations of solutions to the classical Yang-Baxter equation \cite{dYB}. 
In the case of the $\eta$-deformation, the spinning string ansatz was shown to lead to a deformation of the Neumann system in \cite{AM}, where both the deformations 
of the Lax connection and the Uhlenbeck integrals of motion were computed. More recently, the $\eta$-deformation of the complete 
Neumann-Rosochatius system has been found in \cite{AHM}, and the problem of geodesic motion on the $\eta$-deformed two-sphere has been shown 
to be superintegrable \cite{AHMsm}. In this article, we will continue the analysis of the $\eta$-deformed Neumann-Rosochatius system by constructing a general 
set of solutions by integration of the problem in terms of Jacobian elliptic functions (additional solutions corresponding to various string configurations in $\eta$-deformed 
$AdS_5 \times S^{5}$ have been studied before using diverse approaches in references \cite{ALT}-\cite{R2}). 

The remaining part of the article is organised as follows. In section~2 we will introduce the $\eta$-deformation of the Neumann-Rosochatius system. We will consider 
the case of a closed string rotating with two different angular momenta in an $\eta$-deformed three-sphere. In section~3 we will 
employ the Uhlenbeck constants of the system to write the equations of motion of the problem in terms of an elliptic curve. 
We will find a general class of solutions with non-constant radii that can be written in terms of Jacobi elliptic functions. 
We will study the reduction of these elliptic solutions for some limiting values of the deformation parameter of the system. 
We conclude in section~4 with some remarks on our results and a discussion on some related problems. We include an appendix 
where we recover the solutions that we have obtained for limiting values of the $\eta$-deformation by taking the corresponding limit directly at the Lagrangian.


\section{The $\eta$-deformed Neumann-Rosochatius system}

In this article we will be interested in finding solutions to the $\eta$-deformed Neumann-Rosochatius system. These solutions will correspond 
to closed strings rotating in the $\eta$-deformation of $AdS_5 \times S^5$. But before addressing the general problem in the $\eta$-deformed system 
we will first briefly review the spinning string ansatz in the absence of a deformation. For simplicity we will restrict the dynamics of the string to rotation in a three-sphere. 
We will thus take the ansatz $Y_1=Y_2=0$ and 
$Y_3 + i Y_0 = e^{i w_0 \tau}$, together with 
\be
X_1 + i X_2 = x_1 (\sigma) \, e^{i \varphi_1 (\tau, \sigma)} \ , \quad X_3 + i X_4 = x_2 (\sigma) \, e^{i \varphi_2 (\tau, \sigma)} \ ,
\label{ansatz1}
\ee
where $Y_j$ and $X_k$ are, respectively, the embedding coordinates of $AdS_3$ and $S^3$, and we have chosen
\be
\varphi_i (\tau,\sigma) = \omega_i \tau + \alpha_i(\sigma) \ ,
\label{ansatz2}
\ee
with $i=1,2$. As we are going to study closed string solutions, the radial functions and the angles must satisfy the periodicity conditions
\be
x_i (\sigma + 2 \pi) = x_i (\sigma) \ , \quad \alpha_i (\sigma + 2 \pi) = \alpha_i (\sigma) + 2 \pi m_i \ ,
\ee
where $m_i$ are integer numbers that act as winding numbers. When we enter this ansatz in the Polyakov action in the conformal gauge we find the Lagrangian \cite{NR}
\be
L = \frac {\sqrt{\lambda}}{2 \pi} \Big[  \sum_{i=1}^2 \frac {1}{2} \big[ x_i'^2 + x_i^2 (\a_i'^2 - \omega_i^2 ) \big] 
- \frac {\Lambda}{2} ( x_1^2 + x_2^2 - 1) \Big] \ ,
\label{NRq}
\ee
where the prime stands for derivatives with respect to $\sigma$ and $\Lambda$ is a Lagrange multiplier.
This is the Lagrangian of the Neumann-Rosochatius integrable system, which describes a set of oscillators with a centrifugal barrier constrained to move on a sphere.
The equations of motion are
\be
x''_i = (\alpha_i^{\prime 2} -\omega_i^2 +\Lambda) x_i \ , \quad \alpha '_i = \frac{v_i}{x_i^2} \ ,
\ee
where the $v_i$ are some integration constants and the Virasoro constraints read
\be
\sum_{i=1}^2 \big[ x_i^{\prime 2} + x_i^2 (\alpha_i^{\prime 2} + \omega_i^2) \big] = w_0^2 \ , \quad 
\sum_{i=1}^2{x_i^2 \alpha'_i \omega_i} = 0 \ .
\ee
The energy and the angular momenta $J_{i}$ of the string are 
\be
E = \sqrt{\lambda} w_0 \ , \quad J_i = \sqrt{\lambda} \int{\frac{d\sigma}{2\pi} x_i^2 \omega_i} \ .
\ee
The integrability of the Neumann-Rosochatius system follows from the existence of two integrals of motion $I_1$ and $I_2$ in involution, called the Uhlenbeck constants. 
The Uhlenbeck constants were first found in \cite{Uhlenbeck} for the case of the Neumann model, which corresponds to the choice $v_i=0$. In the case of the 
Neumann-Rosochatius system, for arbitrary values of the constants $v_{i}$, they are given by \cite{NR}
\begin{equation}
I_i = x_i^2 +\sum_{j \neq i}{\frac{1}{\omega_i^2 -\omega_j^2} \left[ (x_i x'_j - x_j x'_i)^2 + v_i^2 \frac{x_j^2}{x_i^2} + v_j^2 \frac{x_i^2}{x_j^2} \right]} \ ,
\end{equation}
It is immediate to check that these integrals are constrained by the relation $I_1 + I_2 = 1$. 

We will now move to the case of a spinning string in the $\eta$-deformation of $AdS_{5} \times S^{5}$. In particular we will 
restrict again our analysis to the motion on the deformed sphere. In this case, the spinning string ansatz (\ref{ansatz1}) and (\ref{ansatz2}) leads to 
the Lagrangian of the $\eta$-deformed Neumann-Rosochatius system \cite{AM},
\ba
L & = & \frac{\sqrt{\lambda}}{2\pi} \left[ \frac{(x_1 x'_2 -x'_1 x_2)^2}{(x_1^2 +x_2^2)[1+ \varkappa ^2 (x_1^2 +x_2^2)x_2^2]} 
+ \frac{x_3^{\prime 2}}{(x_1^2 +x_2^2) [1+\varkappa^2 (x_1^2 +x_2^2)]}  \right. \notag \\
& + & \frac{x_1^2 (\alpha_1^{\prime 2} -\omega_1^2)}{1+\varkappa ^2 (x_1^2 +x_2^2)x_2^2} 
+ x_2^2 (\alpha_2^{\prime 2} -\omega_2^2)+\frac{x_3^2 (\alpha_3^{\prime 2} -\omega_3^2)}{1+\varkappa^2 (x_1^2 +x_2^2)} 
+  \frac{ 2\varkappa \omega_1 x_1 x_2 (x_1 x'_2 -x_2 x'_1)}{1+\varkappa ^2 (x_1^2 +x_2^2)x_2^2} \notag \\
& + & \left. \frac{2\varkappa \omega_3 x_3 x'_3}{1+\varkappa^2 (x_1^2+x_2^2)} - 
\frac{\Lambda}{2} (x_1^2 +x_2^2 +x_3^2 -1) \right] \ , \label{N=3etalagrangian}
\ea
where we have written the deformation parameter in terms of $\varkappa=2\eta/(1-\eta^{2})$. 
It is immediate to write down the complete equations of motion for the radial and angular coordinates coming from this Lagrangian. However 
in this article we will only be interested in the case of a string spinning on an $\eta$-deformed three-sphere. Therefore, 
rather than presenting the general set of equations 
we will focus on how we should perform a consistent reduction to capture the dynamics on the deformed three-sphere. 
We can clarify this by inspecting the equation of motion for $x_3$, which is given by
\begin{equation}
\left[ \frac{x'_3}{(x_1^2+x_2^2) (1+\varkappa^2 (x_1^2+x_2^2))} \right] '
=\Lambda x_3 +\frac{x_3 (\alpha_3^{\prime 2} -\omega_3^2)}{1+\varkappa^2 (x_1^2 +x_2^2)} \ .
\end{equation}
We see that $x_3=0$ is a solution independently of the behaviour of the other two coordinates. 
This means that setting $x_{3}=0$ is a consistent truncation from the $\eta$-deformed five-sphere to an $\eta$-deformed 
three-sphere. 
\footnote{
Note that this is not the only reduction that we can perform to obtain a consistent truncation from $S^{5}_{\eta}$ to $S^{3}_{\eta}$. 
For instance, from the equation of motion for $x_{1}$,
\begin{align*}
\frac{x''_1}{r}&=\varkappa^2 \frac{2(x_1^2 + x_2^2) x'_1 x_2 x'_2 +x_1 x_1^{\prime 2} x_2^2 + 2 x'_1 x_2^3 x'_2 - x_1 x_2^2 x^{\prime 2}_2}{r^2} \notag \\
& - 4 \varkappa \, \omega_1 \frac{x_1 x_2 x'_2}{r^2}+\Lambda x_1+\frac{x_1 (\alpha_{1}^{\prime 2} - \omega_1^2)}{r} \left( 1-\varkappa^2 \frac{x_1^2 x_2^2}{r} \right) \ ,
\end{align*}
with $r=1+\varkappa^2 x_2^2 (x_1^2+x_2^2)$, we conclude that the choice $x_1=0$ provides indeed another possible truncation. 
When we set $x_{1}=0$ the Lagrangian becomes
\[
L=\frac{1}{2} \left[ \frac{x_3^{\prime 2}}{x_2^2 (1+\varkappa^2 x_2^2)} +x_2^ 2 (\alpha_2^{\prime 2} -\omega_2^2) 
+ \frac{x_3^2 (\alpha_3^{\prime 2} -\omega_3^2)}{1+\varkappa^2 x_2^2} \right] +\frac{\Lambda}{2} (x_2^2+x_3^2-1) \ ,
\]
which can be easily seen to be equivalent to the one for the $x_{3}=0$ truncation.
} 
The Lagrangian simplifies to
\be
L = \frac{\sqrt{\lambda}}{2\pi} \left[ \frac{x_1^{\prime 2} + x_2^{\prime 2} + x_1^ 2 (\alpha_1^{\prime 2} -\omega_1^2)}{1+\varkappa^2 x_2^2} 
+ x_2^ 2 (\alpha_2^{\prime 2} - \omega_2^2) 
- \frac{\Lambda}{2} (x_1^2 + x_2^2-1) \right] \ .
\ee
The equations of motion for the radial coordinates are given by
\begin{align}
& \frac{x''_1}{1 + \varkappa^2 x_2^2} + 2 \varkappa^2 \frac{x_1 x_1^{\prime 2}}{(1+\varkappa^2 x_2^2)^2} = \frac{x_1( \alpha_1^{\prime 2} -\omega_1^2)}{1 + \varkappa^2 x_2^2} 
+ \Lambda x_1 \label{x1S3equation} \ , \\
& \frac{x''_2}{1+\varkappa^2 x_2^2} - 2 \varkappa^2 \frac{x_2 x_2^{\prime 2}}{(1 + \varkappa^2 x_2^2)^2} = x_2( \alpha_2^{\prime 2} - \omega_2^2 )
- \varkappa^2 x_2 \frac{x_1^{\prime 2} + x_2^{\prime 2} + x_1^2 (\alpha_1^{\prime 2} - \omega_1^2)}{(1+\varkappa^2 x_2^2)^2} + \Lambda x_1 \ , \label{x2S3equation}
\end{align}
and for the angles we find
\be
\alpha '_1 = \frac{v_1}{x_1^2} (1+\varkappa ^2 x_2^2 (x_1^2 +x_2^2)) \ , \quad 
\alpha '_2 = \frac{v_2}{x_2^2} \ . 
\ee
The Virasoro constraints become
\begin{align}
& \frac{x_1^{\prime 2} + x_2^{\prime 2} + x_1^ 2 (\alpha_1^{\prime 2} + \omega_1^2)}{1+ \varkappa^2 x_2^2} 
+ x_2^ 2 (\alpha_2^{\prime 2} +\omega_2^2) = w_0^{2} \ , \label{Virasoro1}\\
& \frac{x_1^2 \alpha'_1 \omega_1}{1 + \varkappa^2 x_2^2} + x_2^2 \alpha '_2 \omega_2 = 0 \label{Virasoro2} \ ,
\end{align}
and the energy and the angular momenta are given now by
\be
E = \sqrt{\lambda} w_{0} \ , \quad J_1 = \int{\frac{d\sigma}{2\pi} \frac{x_1^2 \omega_1}{1+\varkappa^2 x_2^2}} \ , \quad J_2 = \int{\frac{d\sigma}{2\pi} x_2^2 \omega_2} \ .
\ee

We can prove that integrability remains a symmetry of the system after the~$\eta$-deformation by constructing a deformation $\tilde{I}_{i}$ 
of the Uhlenbeck constants which makes them constants of motion again. To find this deformation we are going to assume that
\be
\tilde{I}_{1}=\frac{1}{\omega_1^2 -\omega_2^2} \left[ f(x_1,x_2) \big[x_1^{\prime 2} + x_2^{\prime 2} \big] 
+ \frac{v_1^{2} x_2^2}{x_1^2} +\frac{v_2^{2} x_1^2}{x_2^2} +h(x_1,x_2) \right] \ ,
\ee
and impose that $\tilde{I}'_1=0$. By doing this we find that
\be
- 2 \varkappa^2 f \left( \frac{x_1 x_1^{\prime 3}}{1+\varkappa^2 x_2^2} +\frac{x_1 x'_1 x_2^{\prime 2}}{1+\varkappa^2 x_2^2} \right) 
+ f' x_1^{\prime 2} +f' x_2^{\prime 2}=0 \ ,
\ee
where we have made use of the equations of motion (\ref{x1S3equation}) and (\ref{x2S3equation}). We can easily integrate this relation to get
\be
f(x_2) = \frac{1}{1+\varkappa^2 x_2^2} \ ,
\ee 
where we have set an overall integration constant to 1. We can proceed in the same way to obtain the function $h$. We finally conclude that~\footnote{
The Uhlenbeck constants were constructed using the Lax representation in \cite{AM}. Some immediate algebra 
shows that the constants in \cite{AM} reduce to the ones we present in here along the $x_{3}=0$ truncation.}
\begin{equation}
\tilde{I}_{1}=\frac{1}{\omega_1^2 - \omega_2^2} \left[ \frac{x_1^{\prime 2} + x_2^{\prime 2} + x_1^2 \omega_1^2}{1+\varkappa^2 x_2^2}
- x_1^2 \omega_2^2 + (1 + \varkappa^2) \frac{v_1^{2} x_2^2}{x_1^2} + \frac{v_2^{2} x_1^2}{x_2^2} \right] \ .
\end{equation}
We can follow an identical reasoning to derive the deformation of the second Uhlenbeck constant, 
which turns to verify the extended closure relation $\tilde{I}_1 + \tilde{I}_2=1$.


\section{Spinning string solutions}

We will now focus on the construction of general solutions to the $\eta$-deformed Neumann-Rosochatius system corresponding 
to non-constant radii strings rotating in $S^{3}_{\eta}$. We will first introduce an ellipsoidal coordinate \cite{BT}, defined as the root of the equation 
\begin{equation}
\frac{x_1^2}{\zeta - \omega_1^2} + \frac{x_2^2}{\zeta - \omega_2^2} =0 \ . 
\end{equation}
If we assume that $\omega_1 <  \omega_2$, then the ellipsoidal coordinate will vary from $\omega_1^2$ to $\omega_2^2$. 
When we replace the radial coordinates by the ellipsoidal one 
in the equations of motion we are left with a second-order differential equation for $\zeta$. But we can more conveniently reduce the problem 
to the study of a first-order equation 
by writing the Uhlenbeck constant in terms of the ellipsoidal coordinate \cite{NR}. We find that
\be
\zeta^{\prime 2} = - 4 P_4 (\zeta) \ ,
\label{zetaequation}
\ee
where $P_4(\zeta)$ is the fourth-order polynomial
\begin{align}
P_4 (\zeta) & = - \frac{\varkappa^2 \omega_2^2}{(\omega_1^2-\omega_2^2)^2} (\zeta -\omega_1^2)^2 (\zeta-\omega_2^2)^2 
+ \big( \omega_1^2-(1+\varkappa^2) \omega_2^2+\varkappa^2 \zeta \big) \Big[ \tilde{I}_{1} (\zeta -\omega_1^2) (\zeta-\omega_2^2) \nonumber \\
& + \frac{(1+\varkappa^2) v_1^2}{\omega_1^2-\omega_2^2} (\zeta-\omega_2^2)^2 
+ \frac{v_2^2}{\omega_1^2-\omega_2^2} (\zeta-\omega_1^2)^2 \Big]+(\zeta -\omega_1^2)^2 (\zeta -\omega_2^2) \nonumber \\
&= - \frac {\varkappa^2 \omega_2^2}{(\omega_1^2-\omega_2^2)^2} \prod_{i=1}^4{(\zeta-\zeta_i )} \ . \label{P4definition}
\end{align}
We can solve this equation if  we change variables to
\begin{equation}
\eta^2=\frac{\zeta -\zeta_4}{\zeta_3 -\zeta_4} \ ,
\end{equation}
which transforms equation (\ref{zetaequation}) into
\begin{equation}
\eta^{\prime 2}= - \frac{\varkappa^ 2 \omega_2^2 \zeta_{34}^2}{(\omega_1^ 2 - \omega_2^2)^2} (1- \eta^2) (\eta^2 -\eta_1^2) (\eta^2 -\eta_2^2) \ ,
\end{equation}
where we have defined $\zeta_{ij}=\zeta_i-\zeta_j$ and $\eta_i^2=\zeta_{i4}/\zeta_{34}$. The solution to this equation can be written in terms of the Jacobi elliptic sine,
\be
\eta (\sigma) = \frac{-i \text{ sn}\left[ \pm i \varkappa \omega_2 \zeta_{34} \, \eta_1 \sqrt{(1-\eta_2^2)} (\sigma -\sigma_0) \big/ (\omega_1^2 -\omega_2^2), \nu \right]}{\sqrt{1-\frac{1}{\eta_2^2} - \text{ sn}^2 
\left[ \pm i \varkappa \omega_2 \zeta_{34} \, \eta_1 \sqrt{(1-\eta_2^2)} (\sigma -\sigma_0)/(\omega_1^2 -\omega_2^2) , \nu \right]}} \ ,
\ee
where the elliptic modulus is given by
\be
\nu = \frac{(1-\eta_1^2) \eta_2^2}{(1-\eta_2^2) \eta_1^2} = \frac{\zeta_{13} \zeta_{24}}{\zeta_{14} \zeta_{23}} \ ,
\ee
and $\sigma_0$ is an integration constant that we can set to zero by performing a rotation. Therefore we conclude that
\be
x_1^2 (\sigma) =\frac{\omega_1^2 -\zeta_4}{\omega_1^2 -\omega_2^2} - \frac{\zeta_{34}}{\omega_1^2 - \omega_2^2} \frac{ \zeta_{24} 
\text{ sn}^2\left[ \pm \varkappa \omega_2 \sqrt{\zeta_{14} \zeta_{23}} \sigma / (\omega_1^2 -\omega_2^2) , \nu \right]}{  \zeta_{23} + \zeta_{24}  
\text{ sn}^2 \left[ \pm \varkappa \omega_2 \sqrt{\zeta_{14} \zeta_{23}} \sigma / (\omega_1^2 -\omega_2^2) , \nu \right]} \ . 
\label{ellipticr1} 
\ee
Now we could use this expression to write the energy as a function of the winding numbers and the angular momenta. However, the first step in this direction, 
which is finding the winding numbers and the momenta in terms of the integration constants $v_i$ and the angular frequencies $\omega_i$, 
already leads to complicated integrals. Instead of following this path, which leads to cumbersome and non-illuminating expressions, in what follows we will analyse 
the problem in some interesting regimes of the deformation parameter. Upon inspection of the polynomial (\ref{P4definition}) it is clear that there are two limits that simplify the evaluation of the roots, namely 
$\varkappa = \infty$ and $\varkappa = i$.~\footnote{From an algebraic point of view, the $\varkappa = i$ limit behaves in the same way as the limit 
of pure NS-NS flux in the analysis of the deformation by flux of the Neumann-Rosochatius system \cite{HN}.} The fate of the deformed ten-dimensional 
background in each of these limits has been studied in reference~\cite{HRT}. In the case where $\varkappa = \infty$ the deformed 
ten-dimensional metric is T-dual to de Sitter space times the hyperboloid, $dS_5 \times H^5$, which can also be understood as a flipped double Wick rotation of $AdS_{5} \times S^{5}$. 
On the other hand, in the limit of imaginary deformation, $\varkappa = i$, the deformed ten-dimensional metric turns into a pp-wave type background. 
  
We will study first the case where $\varkappa = \infty$. To analyze this limit we will consider two different possible choices of our physical parameters. 
We will first set $v_2=\omega_1=0$. With this choice, the roots of (\ref{P4definition}) become 
\begin{align}
\zeta_{1,2} & = \frac {1}{2} \Big[ \tilde{I}_1 \omega_2^2-v_1^2 (1+\varkappa^2) 
\pm \sqrt{	\big[\tilde{I}_1 \omega_2^2-v_1^2 (1+\varkappa^2) \big]^2+4 v_1^2 \omega_2^2(1+\varkappa^2)} \, \Big] \ , \nonumber \\ 
\zeta_3 &=\frac{\omega_2^2 (1+\varkappa^2)}{\varkappa^2} \ , \quad \zeta_4 = \omega_2^2 \ .
\end{align}
Therefore in the $\varkappa = \infty$ limit one of the roots goes to (minus) infinity and the degree of the polynomial reduces to three. 
In order to take the limit at the level of the solution, we need to make sure that we can send $\zeta_1$ to minus infinity in a controlled way. This requires writing 
equation~(\ref{ellipticr1}) in the form
\be
x_1^2 (\sigma) =\frac{\omega_1^2 -\zeta_4}{\omega_1^2 -\omega_2^2} - \frac{\zeta_{14}}{\omega_1^2 - \omega_2^2} \frac{ \zeta_{24} 
\text{ sn}^2 \left[ \pm \varkappa \omega_2 \sqrt{\zeta_{34} \zeta_{21}} \sigma / (\omega_1^2 -\omega_2^2) , \nu/(\nu -1) \right]}{  \zeta_{13} + \zeta_{34}  
\text{ sn}^2 \left[ \pm \varkappa \omega_2 \sqrt{\zeta_{24} \zeta_{13}} \sigma / (\omega_1^2 -\omega_2^2) , \nu/(\nu -1) \right]} \ . 
\label{ellipticr1prime}
\ee
After some manipulations we conclude that
\be
x_1^2(\sigma)=\frac{\zeta_4}{\omega_2^2} - \frac{\zeta_{34}}{\omega_2^2} \text{ sn}^2 \left[ \mp \varkappa \sqrt{\zeta_{24} \zeta_{21}} \, \sigma / \omega_{2} \, , 
\zeta_{34}/\zeta_{24}\right] + \dots  \ , 
\ee
which when we enter explicitly the remaining roots becomes 
\be
x_1^2(\sigma) = 1 + \frac{1}{\varkappa^2} \text{ sn}^2 \left[ \mp \varkappa \sqrt{- \omega_2^2 \tilde{I}_2} \, \sigma , 
- v_1^2 / \omega_2^2 \tilde{I}_2 \right] + \dots \ ,
\label{gensolinfty}
\ee 
where we have made use of the closure of the $\eta$-deformed Uhlenbeck constants, $\tilde{I}_{1} + \tilde{I}_{2}=1$. In the appendix we will study this 
solution in some detail, and find an expansion for its energy in terms of the angular momentum. 

The second interesting choice of parameters in the $\varkappa = \infty$ limit is~$\omega_2=v_1=v_2=0$. In this case the roots become
\be	
\zeta_1 = - \infty \ , \quad \zeta_2 = 0 \ , \quad \zeta_3 = \frac{\omega_1^2 \tilde{I}_2}{1+\varkappa^2 \tilde{I}_1} \ ,  \quad \zeta_4 = \omega_1^2 \ ,
\label{roots2}
\ee
and the degree of the polynomial is again reduced to three. The solution is given by~\footnote{A similar result was obtained for the pulsating string ansatz in reference \cite{BP2}.} 
\be
x_2^{2}(\sigma) = \frac{\zeta_2}{\omega_1^2} - \frac{\zeta_{24}}{\omega_1^2} \text{ sn}^2 \left[ \sqrt{\zeta_{23} (1+\varkappa^2 \tilde{I}_1 )} \, \sigma , 
\zeta_{24} / \zeta_{23} \right] \ ,
\ee
which when we enter the roots (\ref{roots2}) turns into
\be
x_2^{2}(\sigma) = \frac{\tilde{I}_2}{1+\varkappa^2 \tilde{I}_1} \text{ sn}^2 \left[ \sqrt{-\omega_1^2 (1+\varkappa^2 \tilde{I}_1)} \, \sigma , 
\frac{\tilde{I}_2}{1+\varkappa^2 \tilde{I}_1} \right] \ , \label{omega1notzero}
\ee
where we have made use of the relation $ \hbox{sn}(u,m) =\hbox{sn}(\sqrt{m}u,\frac{1}{m})/ \sqrt{m}$. We must note that this solution contains four different regimes. 
In the cases where either $\tilde{I}_1 \geq 1$ or $\tilde{I}_1 < -1/\varkappa^2$ 
we have to analytically continue the $x_2$ coordinate to $ix_2$. On the contrary, the region where $0<\tilde{I}_1 <1$ requires the continuation 
of the $x_1$ coordinate instead. Finally, in the region with $-1/\varkappa^2 < \tilde{I_1} <0$ we are left with a circular solution, 
which completely disappears in the limit $\varkappa = \infty$, where the sphere gets deformed into the hyperboloid \cite{HRT}.

We will now move to the study of the $\varkappa = i$ limit. In this limit the contribution from the constant $v_1$ is negligible and $\omega_1$ 
reduces to a shift in the Uhlenbeck constant. 
Therefore $\omega_2$ and $v_2$ are the only relevant free parameters. Again, we will consider two different cases. We will first set $v_{2}=0$. 
In this case the roots of (\ref{P4definition}) behave like
\be
\zeta_{1,2} = 0 + \cdots \ , \quad \zeta_3 = \tilde{I}_1 \omega_2^2 \ , \quad \zeta_4 =\omega_2^2 \ .
\ee
Upon substitution and after some immediate algebra we find
\be
x_1^2(\sigma)= \frac{\tilde{I}_1}{1 + \tilde{I}_2 \cosh^2 \Big( \sqrt{\omega_2^2 \tilde{I}_1} \sigma \Big)} \ .
\label{omega2notzero}
\ee
We will next consider the case with $\omega_2=0$, where the roots are given by
\be
\zeta_1 \simeq \zeta_2= \omega_1^2 + \cdots \ , \quad 
\zeta_3 = \frac{v_2^2 \omega_1^2}{v_2^2 - \omega_1^2 \tilde{I}_2} \ , \quad \zeta_4=-\infty \ ,
\ee
where we have kept $\omega_{1} \neq 0$ to avoid the need to redefine $\tilde{I}_{1}$. 
After some immediate algebra we conclude that
\be
x_1^2 (\sigma) = \frac{\omega_1^2 \tilde{I}_2 }{v_2^2 - \omega_1^2 \tilde{I}_2} \sech^2 \big( \tilde{I}_2  \sigma \big) \label{omega2zero}\ .
\ee
Now we can eliminate the $\omega_1$ factor by the redefinition $v_2=\tilde{v}_2 \omega_1$. As we have noted above, this is a consequence 
of the fact that the term encoding the dependence on $\omega_1^2$ in the Uhlenbeck integral becomes a constant in the $\varkappa=i$ limit, 
making it a dummy variable.

We must point out that although the solutions for $v_2=0$ and $\omega_2=0$ seem completely different, they are deeply related. 
This relation is not explicit from point of view of the Uhlenbeck constant, but it will be evident once we have written the Lagrangian associated to each limit. 
We will explore this connection in the Appendix.

To conclude this section, we will consider the case where the radii are taken to be constant, which allows to obtain the energy of the string as an expansion in the 't Hooft coupling and the angular momentum for 
arbitrary values of the deformation parameter. When we set to zero the derivatives in the equations of motion 
and solve for the Lagrange multiplier we find that
\be
\frac{\alpha_1^{\prime 2}-\omega_1^2}{1+\varkappa^2 x_2^2} = \alpha_2^{\prime 2} - \omega_2^2 - \varkappa^2 \, \frac{x_1^ 2 (\alpha_1^{\prime 2} - \omega_1^2)}{(1+\varkappa^2 x_2^2)^2} \ .
\ee
We can rewrite this expression as
\begin{equation}
(1 + \varkappa^2 x_2^2)^{2} = (1+\varkappa^2) \frac{m_1^2 -\omega_1^2}{m_2^2 -\omega_2^2} \ ,
\label{constantr}
\end{equation}
where we have used the constraint $x_1^2 + x_2^2=1$ and the fact that $m_i=\alpha'_i$ because the angular velocities are constant when the radii are constant. 
From this relation it is immediate to conclude that in the limit $\varkappa = i$ the solution reduces to $x_1=0$ and $x_2=1$, together with either zero winding number 
or zero total angular momentum because of the Virasoro constraint~(\ref{Virasoro2}). However solving 
equation (\ref{constantr}) exactly for arbitrary values of the deformation together with the Virasoro constraint leads to an algebraic equation of sixth degree. 
Instead of trying to solve the problem directly, we can write the solution as a power series expansion in inverse powers of the total angular momentum. 
We get \footnote{There is an additional possible expansion, depending on the choice of signs for the winding numbers.}
\begin{align}
x_1^2 & = \frac{km_2}{k m_2-m_1} + \frac {\lambda}{2 J^{2}} 
\frac{k m_1 m_2 (m_1 +m_2) (m_1 -m_2)^3 (m_1^2 - 2 k m_1 m_2 +m_2^2)}{(k m_1 -m_2)^2 (m_1 -k m_2)^4} + \cdots \ , \\
x_2^2 &= \! \frac{m_1}{m_1-km_2} - \frac {\lambda}{2 J^{2}} 
\frac{k m_1 m_2 (m_1+m_2) (m_1 -m_2)^3 (m_1^2 -2 k m_1 m_2 +m_2^2)}{(k m_1 -m_2)^2 (m_1 -k m_2)^4} + \cdots \ , 
\end{align}
for the radial coordinates, and 
\begin{align}
\omega_1 & = \frac{J}{\sqrt{\lambda}} \frac{k m_1 \! - \! m_2}{m_1 \! - \! m_2} 
+ \frac {\sqrt{\lambda}}{2 J} \frac{k m_1 (m_1 \! + \! m_2) (m_1 \! - \! m_2)^2 (m_1^2 -2 k m_1 m_2 +m_2^2)}{(km_1 -m_2)^2 (m_1 -km_2)^2} + \cdots \ , \\
\omega_2 & = \frac{J}{\sqrt{\lambda}} \frac{m_1 \!  - \! k m_2}{m_1 \! - \! m_2} 
+ \frac {\sqrt{\lambda}}{2 J} \frac{k m_2 (m_1 \! + \! m_2) (m_1 \! - \! m_2)^2 (m_1^2 -2 k m_1 m_2 +m_2^2)}{(km_1 -m_2)^2 (m_1 -km_2)^2} + \cdots \ ,
\end{align}
for the angular frequencies, where for convenience we have defined $k=\sqrt{1+\varkappa^2}$. 
Using now equation~(\ref{Virasoro1}) it is immediate to write the dispersion relation,
\begin{equation}
E^2 = J^2 \frac{(m_1^2 -2 k m_1 m_2 +m_2^2)}{(m_1-m_2)^2} + \lambda \frac{m_1 m_2 (m_1^2 -2km_1 m_2 +m_2^2)}{(km_1 -m_2) (km_2 -m_1)} + \dots \ .
\end{equation}
In the absence of deformation, this expression reduces to the expansion for the energy of a circular string rotating in a three-sphere  
with two different angular momenta \cite{NR}. 


\section{Conclusions}

In this article we have constructed a general class of solutions to the $\eta$-deformed Neumann-Rosochatius system. 
The solutions that we have found correspond to closed strings with non-constant radii rotating with two different angular momenta 
in an $\eta$-deformed three-sphere. The solutions can be written in terms of Jacobian elliptic functions. We have studied the problem 
for some limiting values of the $\eta$-deformation, which allow to reduce the degree of the polynomial of the elliptic surface. In particular, 
we have considered the limit $\eta=1$, where the deformed $AdS_{5} \times S^{5}$ target space becomes $dS_{5} \times H^{5}$, 
and the limit where $\eta=i$, which corresponds to a string moving in a pp-wave type background. We have also solved the case of strings 
with constant radii as an expansion in the 't Hooft coupling and the total angular momentum for arbitrary values of the deformation parameter. 

There are several interesting directions that can be followed to extend our analysis. An immediate one is the study of closed strings spinning 
in the complete $\eta$-deformed five-sphere, or in the complete background. The hyperelliptic curve solving the problem in this higher dimensional case 
should again get reduced for the $\eta=1$ and $\eta=i$ limiting values. A more appealing problem comes from the relation between different integrable deformations 
of $AdS_{5} \times S^{5}$. In fact, the $\eta$-deformation that we have considered in this article can be recovered from the $\lambda$-deformation 
by performing a scaling limit together with an analytical continuation of the coordinates when the deformation parameters are adequately identified \cite{HRT,Vicedo}. 
It would be worthwhile to investigate the meaning of our solutions and of the $\eta$-deformation of the Neumann-Rosochatius system from the point of view 
of the $\lambda$-deformation (see reference \cite{R3} for a related recent discussion on this point). 


\newpage

\vspace{8mm}

\centerline{\bf Acknowledgments}

\vspace{2mm}

\no
The work of R.~H. is supported by grant FPA2014-54154-P and by BSCH-UCM through grant GR3/14-A 910770. 
We thank D.~Medina-Rinc\'on for discussions and S.~J.~van Tongeren for comments on the manuscript. 
J.~M.~N. also thanks the Institut de Physique Th\'eorique at CEA Saclay 
for hospitality and financial support during part of this work. The research leading to these results has received funding 
from the European Community's Seventh Framework Programme FP7/2007-2013 under grand agreement No 317089 (GATIS) 
and by CT45/15 grant from the Universidad Complutense of Madrid.


\appendix

\renewcommand{\theequation}{\thesection.\arabic{equation}}
\csname @addtoreset\endcsname{equation}{section}


\section{Appendix}

In this Appendix we will analyse the solutions that we have constructed in this article in the cases where $\varkappa = \infty$ and $\varkappa = i$ 
by performing the corresponding limit directly at the level of the Lagrangian. 
In order to deal with this problem it will be useful to think of the change of variables that brings the kinetic term in the deformed Lagrangian 
to canonical form, which is given by $x_2=\text{sn} \left( \phi , -\varkappa^2 \right)$. In the variable $\phi$ the Lagrangian turns into
\begin{equation}
L \! = \! \frac{1}{2} \!  \left[ \phi^{\prime 2} - \omega_2^2 \text{sn}^2 \left( \phi , -\varkappa^2 \right) \! - \! \frac{v_2^2}{\text{sn}^2 \left( \phi , -\varkappa^2 \right)} 
\! - \! \frac{\omega_1^2 \left( 1 + \frac{1}{\varkappa^2} \right)}{1+\varkappa^2 \text{sn}^2 \left( \phi , -\varkappa^2 \right)} -\frac{(1+\varkappa^2) v_1^2}{\text{cn}^2 \left( \phi , -\varkappa^2 \right)}\right]  .
\end{equation}
In the limit $\varkappa = i$ the change of variables reduces to $x_2=\tanh \phi$, together with $x_1=\sech \phi$, and thus the Lagrangian becomes
\be
L_{i} = \frac{1}{2} \left[ \phi^{\prime 2} - \frac{v_2^2}{\sinh^2 \phi} - \frac{\omega_2^2}{\cosh^2 \phi} \right] \ ,
\ee
where we have shifted the Lagrangian by a constant to rewrite the term associated with~$v_2^2$ with a hyperbolic secant instead of a hyperbolic cotangent. To find the limit $\varkappa = \infty$ 
we need to transform the elliptic sine, because its fundamental domain is defined when the elliptic modulus is contained between $0$ and $1$. We will write
\begin{equation}
\text{sn} (\phi, -\varkappa^2)=\frac{\text{sd} 
\left(\sqrt{1+\varkappa^2} \phi , \frac{\varkappa^2}{1+\varkappa^2} \right)}{\sqrt{1+\varkappa^2}} \simeq \frac{\sinh (\varkappa \phi)}{\varkappa} \ .
\end{equation}
Therefore the change of variables is given by $\varkappa \, x_2=\sinh \varkappa \phi=\sinh \tilde{\phi}$, which leads to 
\footnote{The extra term $(v_1^2 +\varkappa^2 v_1^2)$ accompanying $\omega_2^2$ comes from the expansion of the Jacobi cosine. 
Also, although not obvious, taking this limit implicitly assumes $x_2 \ll \mathcal{O} (\varkappa^{-1})$. That is the reason why the $1-x_2^2$ factor dividing the kinetic term 
disappears, in spite by direct substitution of the change of variables it is subleading in $\varkappa^{-2}$.}
\be
L_{\infty} = \frac{1}{2\varkappa^2} \left[ \tilde{\phi}^{\prime 2}-\left( \omega_2^2 +v_1^2 +\varkappa^2 v_1^2 \right) 
\sinh^2 \tilde{\phi} - \frac{\varkappa^4 v_2^2}{\sinh^2 \tilde{\phi}} - \frac{(1+\varkappa^2 ) \omega_1^2}{\cosh^2 \tilde{\phi}} \right] \ . 
\ee
Both limits lead thus to the same kind of Lagrangian, although with different coefficients in front of the potential terms. In what follows 
we will treat both of them simultaneously. However, even in these limiting cases the Lagrangian is not easy to handle unless 
some additional simplifications are performed. 
These simplifications will come from various convenient choices of the physical parameters entering the problem. 
We will start by considering the easiest choice of parameters on the Lagrangian, which is that where only the potential 
with the square of the hyperbolic sine survives. Then
\begin{equation}
L = \frac {1}{2 \varkappa^2} \Big[ \phi'^2 - \alpha^2 \sinh^2 \phi \Big] \ ,
\label{Lreduced}
\end{equation}
with $\alpha$ a constant which will depend on which of the two limits we are taking. The equation of motion is then
\begin{equation}
\phi'' = - \alpha \sinh \phi \cosh \phi \ ,
\end{equation}
and can be solved in terms of the Jacobi amplitude,
\be
\phi (\sigma) =\pm i \text{ am}\left(\sqrt{ \alpha^{2}+c } \ \sigma,\frac{\alpha^{2}}{\alpha^{2}+c} \right) \ ,
\ee
where $c$ is a constant that has to be fixed by imposing periodicity of $x_i$ (we have made use again of our freedom in the choice of $\sigma$ 
to eliminate an additional integration constant). Note that in general, depending on the sign of $\alpha^2+c$, we find two different solutions. 

We will now focus on the limit $\varkappa = \infty$. In this case the solutions are given by
\begin{align}
x_2^{2} (\sigma) &= - \frac{1}{\varkappa^2} \text{sn}^{2} \left(\sqrt{\alpha^2+c} \, \sigma ,\frac{\alpha^2}{\alpha^2+c} \right) \ , \quad \text{when} \ \alpha^2 + c >0 \ , \label{sol1} \\
x_2^{2}  (\sigma) & = \frac{1}{\varkappa^2} \text{sc}^{2} \left(\sqrt{-(\alpha^2+c)} \, \sigma ,\frac{c}{\alpha^2 + c} \right) \ ,
\quad \text{when} \ \alpha^2 + c< 0 \ .
\end{align}
Note that in both cases we have to analytically continue to hyperbolic space. This is in agreement with the results obtained in~\cite{HRT}, 
where in the limit  $\varkappa = \infty$ the deformed sphere becomes a hyperboloid.
We must however stress that the periodicity condition for each solution is different. 
This is because the real periodicity of the sn$^2$ function is given by $2\text{K}(m)$ while 
its imaginary periodicity is $2i\text{K}(1-m)$, where $\hbox{K}(x)$ is the complete elliptic integral of first kind. 
Furthermore, the presence of the Jacobi $\hbox{sc}(u,m)$ function in the case where $\alpha^{2}+c <0$ leads to a divergence when evaluating the angular momentum,  
so from now on we will only consider relation (\ref{sol1}). This case corresponds to solution (\ref{gensolinfty}) 
once we set $\alpha^2+c = \varkappa^2 \omega_2^2(\tilde{I}_1-1)$.
The periodicity condition implies
\be
\frac{n}{\pi} \text{K}\left( \frac{\alpha^2}{\alpha^2+c} \right)=\sqrt{\alpha^2+c} \ ,
\ee
which in general has no analytical solution. However, as $\alpha^{2}$ 
grows like $\varkappa^2$ we can assume that $c/\alpha^2$ is small enough to perform a series expansion in both sides of the equality. 
Then if we recall now that
\be
\hbox{K}[1-x] \simeq -\frac{\log (x)}{2}+2\log (2) \ ,
\ee
we find  
\begin{equation}
c \simeq \frac{n \alpha}{\pi} \, \hbox{W} \big( 16 \alpha \pi \, e^{-2 \alpha \pi /n}/n \big) \ ,
\label{LambertW}
\end{equation}
where $\hbox{W}(x)$ is the Lambert W function. In fact, it is easy to check that our assumption becomes true very fast, because 
when $n=10$ and $\alpha^2=200$ we already have $c/\alpha^2 \approx 0.0014$. Now, as we need to set $v_2=\omega_1=0$ 
to bring the Lagrangian to the form (\ref{Lreduced}), we have $J_1=m_2=0$ and therefore we only need to compute the angular momentum, 
\be
J_2 = \int{\frac{d\sigma}{2\pi} x_2^2 \omega_2} = 
\frac{\omega_2}{\varkappa^2} \ \frac{\alpha^2+c}{\alpha^2} \left[ 1- \frac {\hbox{E} \left( \frac{\alpha^2}{\alpha^2+c} \right)}{\hbox{K} \left( \frac{\alpha^2}{\alpha^2+c} \right)}  \right] \ , 
\ee
and the winding number, 
\be
m_1 \! = \! \int{\frac{d\sigma}{2\pi} \, \frac{v_1 (1-\varkappa^2 x_2^2)}{x_1^2} } 
= v_1 \left[ \frac{1+\varkappa^2}{\varkappa^2} \ \frac{\Pi \! \left( \! -\frac{1}{\varkappa^2} , \frac{\alpha^2}{\alpha^2+c} \right)}{\hbox{K} \left( \frac{\alpha^2}{\alpha^2+c} \right)} -1 \right] \ ,
\ee
where we have used the periodicity condition to simplify both expressions, and $\hbox{E}(x)$ and $\Pi(n,x)$ are, respectively, the complete elliptic integrals of the second and the third kind. 
If we take now the large $\alpha^2$ limit, we conclude that
\begin{equation}
J_2 = J = \frac{\omega_2}{\varkappa^2} + \cdots \ ,
\end{equation}
where we have used that the first elliptic integral diverges at $x=1$, while the second elliptic integral goes to 1. The winding number can also be expanded as
\begin{align}
m_1=\frac{3 v_1}{2\varkappa^2} + \cdots \ .
\end{align}
The only thing left is to find the dispersion relation,
\begin{align}
E^2 & = \int{\frac{d\sigma}{2\pi} \left( \frac{x_1^{\prime 2} - x_2^{\prime 2}}{1-\varkappa^2 x_2^2} +\frac{v_1^2 (1 - \varkappa^2 x_2^2)}{x_1^2} - x_2^2 \omega_2^2 \right)} \nonumber \\
& = \alpha^2 \left[ 1 -\frac{(1+\varkappa^2 )\alpha^2+c}{\varkappa^2 \alpha^2} \ \frac{\Pi \left( -1, \frac{\alpha^2}{\alpha^2+c} \right)}{\hbox{K} \left( \frac{\alpha^2}{\alpha^2+c} \right)} \right] + m_1 v_1 - J \omega_2  \ ,
\end{align}
that can be easily expanded to find
\be
E^2 = - \frac{\alpha^2}{2\varkappa^2}+m_1 v_1 -J\omega_2 + \cdots =-\frac{2 \varkappa^4 m_1^2}{9} + \frac{4 \varkappa^2 m_1^2}{9} - \frac {3 \varkappa^2 J^2}{2} + \cdots \ .
\ee
Note that the energy that we have obtained is purely imaginary. This is a consequence of the fact that 
the $\eta$-deformed Anti-de Sitter factor in the metric reduces to de Sitter space in the $\varkappa\rightarrow \infty$ limit \cite{HRT}. Therefore the time coordinate is analytically continued, 
which explains the negative sign on the square of the energy.

We will next move to the choice of parameters that brings the Lagrangian to the form
\begin{equation}
L = \frac {1}{2 \varkappa^2} \Big[ \phi'^2 -\frac{\alpha^2}{\cosh^2 \phi} \Big] \ .
\label{L2}
\end{equation}
Instead of writing the equations of motion for this Lagrangian and trying to integrate them it is more convenient to write the corresponding Hamiltonian, 
\be
H = \phi'^2 +\frac{\alpha^2}{\cosh^2 \phi} \ ,
\ee
and make use that is is a conserved quantity to find $\phi$ from direct integration.  We conclude that 
\begin{equation}
\arcsinh \phi=\sqrt{ \frac{|\alpha^2-H|}{H}} \sinh (\sqrt{H} \sigma) = \varkappa x_2 \ .
\label{s1}
\end{equation}
However this solution is not the same as the one we obtained by analysing the roots of the quartic polynomial (\ref{omega1notzero}).  
The reason for this mismatch is that, as we have previously discussed, the Lagrangian 
that we have written implicitly ignores the $1/(1-x_2^2)$ term in the kinetic energy as it is subleading in $\varkappa$. 
If we restore this factor the Hamiltonian reads
\begin{equation}
\bar{H}=\frac{\varkappa^2 \phi'^2}{\varkappa^2 - \sinh^2 \phi} +\frac{\alpha^2}{\cosh^2 \phi} \ ,
\end{equation}
which can be integrated to obtain
\begin{align}
\varkappa x_2 & =\arcsinh \phi = \pm \varkappa \, \text{sn} \Big( \sqrt{\frac{\bar{H}- \alpha^2}{\varkappa^2}} \sigma , \frac{\bar{H} \varkappa^2}{\alpha^2 - \bar{H}} \Big) \notag \\
&=  \sqrt{\frac{\alpha^2 - \bar{H}}{\bar{H}}} \text{sn} \left( \sqrt{-\bar{H}} \sigma , \frac{\alpha^2 - \bar{H}}{\bar{H} \varkappa^2} \right) \ .
\label{s2}
\end{align}
In fact, solution (\ref{s1}) can be recovered from (\ref{s2}) by taking the $\varkappa = \infty$ limit on the last expression. 
From the point of view of equation (\ref{omega1notzero}), ignoring the $1/(1-x_2^2)$ term in the kinetic energy can be understood as taking explicitly 
the limit $\zeta_3 \ll \zeta_4$, which implies
\be
\zeta (\sigma) = \zeta_3 \sinh^2 \Big[ \sqrt{\omega_1^2 (1+\varkappa^2 \tilde{I}_1)} \, \sigma \Big] \ .
\ee
We can match the solutions obtained from the Uhlenbeck constants to the solutions obtained 
from the equation of motion by identifying $\bar{H}=\omega_1^2 (1+\varkappa^2 \tilde{I}_1)$ and $\alpha^2=(1+\varkappa^2) \omega_1^2$.

We can now easily find the solution to the equations of motion for the $\varkappa = i$ limit in the case where $v_2=0$.  
To do that we only have to use the transformation $x_1=\sech \phi$ in solution~(\ref{s1}) and recall that $\sech[ \arcsinh (x)]=1/\sqrt{1+x^2}$. We find that
\be
x_1^2 (\sigma) =\frac{H}{H-|\alpha^2-H| \cosh^2 ( \sqrt{H} \sigma )} \ .
\ee
This solution can be identified with equation (\ref{omega2notzero}) 
once we perform the substitutions $\alpha^2=\omega_2^2$ and $H=\tilde{I}_1 \omega_2^2$. Note that for the formula to have the correct sign we need $\tilde{I}_1 \leq 1$, 
which is equivalent to the condition $\omega_1^2 \leq \zeta_i \leq \omega_2^2$ on the roots of the elliptic curve.

To conclude our discussion we will address a third simplification of the Lagrangian, 
\be
L = \frac {1}{2 \varkappa^2} \Big[ \phi'^2 -\frac{\beta^2}{\sinh^2 \phi} \Big] \ .
\label{L3}
\ee
We can get all possible solutions to this Lagrangian from the solutions of Lagrangian~(\ref{L2}) after substituting $\phi \rightarrow \phi +\frac{i\pi}{2}$ and $\beta^2 \rightarrow -\alpha^2$. 
Let us examine one of the solutions, corresponding to the limit $\varkappa=i$ and $\omega_2=0$. If we choose the solution with the hyperbolic cosine we get
\be
x_1^{2}(\sigma) = \sech^{2} \left( \phi \pm\frac{i\pi}{2} \right) = \frac{H}{-\beta^2-H} \sech^{2} (\sqrt{H} \sigma ) \ ,
\ee
which is equivalent to solution (\ref{omega2zero}) once we identify $\beta^2=-v^2_2$ and $H=\tilde{I}_2$.


\end{document}